\documentclass[prb,twocolumn,amsmath,amssymb,superscriptaddress]{revtex4}
\usepackage[pdftex]{graphicx}
\usepackage[outdir=./]{epstopdf}
\usepackage{enumerate,ulem}
\newcommand \beq{\begin{eqnarray}}
\newcommand \eeq{\end{eqnarray}}

\newcommand{\bm}[1]{\boldsymbol{#1}}

\newcommand{\bfr}{\bm{r}}
\newcommand{\bfk}{\bm{k}}

\newcommand{\bfp}{\bm{p}}
\newcommand{\bfsigma}{\bm{\sigma}}

\newcommand{\bfn}{\bm{n}}

\newcommand{\vf}{v_\mathrm{F}}

\newcommand{\im}{\mathrm{Im}}
\newcommand{\sgn}{\mathrm{sgn}}

\newcommand{\vecv}[2]{\left( \begin{array}{c} #1  \\ #2 \end{array}\right)}
\newcommand{\matr}[4]{\left( \begin{array}{cc} #1 & #2 \\ #3 & #4 \end{array}\right)}

\newcommand{\modify}[1]{{#1}}

\begin{document}

\title{Skyrmion-induced anomalous Hall conductivity on topological insulator surfaces}
\author{Yasufumi Araki}
\affiliation{Institute for Materials Research, Tohoku University, Sendai 980-8577, Japan}
\affiliation{Frontier Research Institute for Interdisciplinary Sciences, Tohoku University, Sendai 980-8578, Japan}
\author{Kentaro Nomura}
\affiliation{Institute for Materials Research, Tohoku University, Sendai 980-8577, Japan}

\begin{abstract}
Electron spin-momentum locking together with background magnetic textures can significantly alter the electron transport properties.
We theoretically investigate the electron transport at the interface between a topological insulator and a magnetic insulator with magnetic skyrmions on the top.
In contrast to the conventional topological Hall effect in normal metals,
the skyrmions yield an additional contribution to the anomalous Hall conductivity even in the absence of in-plane magnetic texture,
arising from the phase factor characteristic to Dirac electrons acquired at skyrmion boundary.
\end{abstract}

\maketitle

\section{Introduction}
Physics of magnetic textures has been a significant topic in recent studies on magnetic materials,
to make use of them as efficient carriers of information
\cite{Zutic_Fabian_DasSarma,Brataas_Bauer_Kelly,Tatara_Kohno_Shibata}.
Skyrmion, in particular, is a particlelike magnetic excitation with a swirling texture in two dimensions (2D),
in which the spin at the core and those at the perimeter point in the opposite directions \cite{Nagaosa_Tokura,Fert}.
Such non-collinear magnetic textures are observed
in non-centrosymmetric magnets, such as helimagnetic conductor $\mathrm{MnSi}$ (Refs.~\onlinecite{Muhlbauer,Tonomura_Tokura})
and magnetic insulator (MI) $\mathrm{Cu}_2 \mathrm{O} \mathrm{Se} \mathrm{O}_3$ (Refs.~\onlinecite{Seki_Tokura,Adams}).
Skyrmions form a periodic lattice, called skyrmion crystal,
with the lattice constant around $10$-$100\mathrm{nm}$.
One of the most important features of skyrmions is the {idea} of emergent electromagnetic fields for conduction electrons,
arising from nonzero Berry curvature in the real space \cite{Nagaosa_Tokura_PhysScr,Nagaosa_Yu_Tokura}.
A conduction electron traveling across the magnetic texture feels the emergent magnetic field,
leading to the so-called topological Hall effect (THE) \cite{Ye_THE,Tatara_Kawamura,Lee_THE,Neubauer_THE}.

In the presence of strong spin-orbit coupling (SOC), on the other hand,
the band topology in the momentum space {significantly affects} the electron transport as well \cite{TKNN}.
A typical example is the surface state of a topological insulator (TI),
where the electrons show 2D linear dispersion with a single band-touching (Dirac) point \cite{Hasan_Kane,Qi_Zhang}.
Under a finite out-of-plane magnetization,
the Dirac point is gapped out and the system acquires a finite Chern number $C=\pm 1$ from the momentum-space Berry curvature,
leading to the well-known quantum anomalous Hall effect (quantum AHE, QAHE) \cite{Qi_Wu_Zhang,Qi_Hughes_Zhang,Yu,Nomura_QAHE,Checkelsky,Chang}.
Since the exchange coupling between the local {in-plane} magnetization and the TI surface electron
can be regarded as the emergent gauge field (vector potential) for the conduction electrons \cite{Garate_Franz},
topological magnetic textures on TI surfaces can give rise to even richer electronic properties:
vortices and domain walls, for instance, are supposed to host zero modes,
leading to electric charging of those textures \cite{Nomura_DW,Tserkovnyak_Loss,Tserkovnyak_Pesin_Loss}.
Recent theoretical studies predict that magnetic skyrmions on TI surface can be charged as well
\cite{Hurst_Galitski,Andrikopoulos}.
Transport measurements in magnetic TI heterostructures discovered the coexistence of
the THE related to the real-space Berry curvature from skyrmions
and the AHE related to momentum-space counterpart from SOC \cite{Yasuda},
while the transport is dominated by the bulk states,
{leaving the surface transport ambiguous}.

In this work, we theoretically examine the electron transport on TI surface in the presence of skyrmions,
which possibly accounts for the interface between a TI and a non-centrosymmetric MI.
\modify{Due to electron spin-momentum locking feature on TI surfaces,
the electron spin does not follow the local magnetic texture even under a strong exchange splitting,
which makes the conventional THE scenario arising from the real-space Berry curvature,
based on the adiabatic approximation, unreliable.}
Hence we fully solve the electron scattering problem by a single skyrmion,
and apply its result to the Boltzmann transport theory to estimate the longitudinal and Hall conductivities.
We find that the skyrmions give a sizable additive contribution to the AHE,
due to the skewness in the electron scattering at a skyrmion.
The origin of this skewness is the phase factor acquired at the skyrmion boundary,
which is characteristic to Dirac electrons and is absent in Schr\"{o}dinger electron systems.
\modify{It arises even in the absence of in-plane magnetic texture},
which is totally different from the conventional THE in normal metals.
In this paper, we take $\hbar=1$ and restore it in the final numerical results.

\section{Single-skyrmion problem}
\modify{Let us set up a heterostructure of a TI (e.g. $\mathrm{Bi}_2 \mathrm{Se}_3$) and a MI (e.g. $\mathrm{Cu}_2 \mathrm{O} \mathrm{Se} \mathrm{O}_3$),
with its interface taken parallel to the layers of the TI
so that the coupling between the TI and the MI should be homogeneous.}
We first place a single skyrmion at the center of the 2D infinite space,
to examine how an incoming electron plane wave gets scattered by the skyrmion.
The electron at the TI-MI interface under the magnetic texture $\boldsymbol{n}(\bfr)$ is described by the Hamiltonian
\begin{align}
\mathcal{H} = \vf (\hat{\bfp}\times\bfsigma)_z - \Delta \bfn(\bfr)\cdot\bfsigma,
\end{align}
where the coefficient $\vf$ is the Fermi velocity, $\hat{\bfp} = -i \boldsymbol{\nabla}$ is the momentum operator,
and $\Delta$ is the spin splitting energy from the exchange interaction between the electron and the magnetization.
\modify{Since the interface is parallel to the layers of the TI,
the spin of the interface electron is helical in the momentum space, i.e. spin-momentum locked} \cite{Silvestrov,Zhang_Kane_Mele}.
\modify{Although the TI surface Hamiltonian in realistic systems exhibit the terms beyond the linear order in the momentum $\hat{\boldsymbol{p}}$,
their effect on the electron scattering process that we are interested in is just a slight modulation of electron spin-momentum locking.
It may shift the scattering amplitude quantitatively,
but would not qualitatively alter the angular profile of the scattering behavior (differential scattering cross section).
Thus we will neglect those higher order terms in our present calculation.}
In the numerical results below, we fix the parameters
$v_\mathrm{F} = 0.5 \times 10^6 \; \mathrm{m/s}$ and $\Delta = 10 \; \mathrm{meV}$,
which are typical values in TI-MI heterostructures,
such as $\mathrm{Bi_2 Se}_3$ and yttrium iron garnet (YIG)
\cite{Liu_Wang,Kubota}.

We here fix the skyrmion texture unaffected by the interface electrons,
assuming that it is stabilized by the Dzyaloshinskii-Moriya interaction (DMI) in the MI.
Here a skyrmion with a vortical magnetic texture is energetically more stable than that with a hedgehog texture \cite{Nagaosa_Tokura,Fert}.
A vortex skyrmion can be parametrized by the cylindrical coordinate $\bfr = (\rho,\phi)$ as
\begin{align}
\bfn(\bfr) =
\begin{pmatrix}
 -\sqrt{1-n_z^2(\rho)} \sin\phi \\
 \sqrt{1-n_z^2(\rho)} \cos\phi \\
 n_z(\rho)
\end{pmatrix},
\end{align}
where $n_z(\rho)$ is a scalar function taking the value between $-1$ at the center and $+1$ at the perimeter.
\modify{Here we should note that the in-plane magnetic texture in this ``vortex skyrmion'' can be removed
by the local U(1) gauge transformation $U(\rho) = \exp[i\frac{\Delta}{v_\mathrm{F}} \int_0^{\rho} d\rho' \sqrt{1-n_z^2(\rho')}]$,
which states that only the sign flip in $n_z(\rho)$ is essential in our calculation.
We thus employ the ``hard-wall'' approximation
$n_z(\rho) = \sgn(\rho - R_\mathrm{S})$,
with $R_\mathrm{S}$ the skyrmion radius} \cite{Hurst_Galitski}.
Such a structure is likely to be realized under a strong out-of-plane magnetic anisotropy.
It is not responsible for the THE in normal metals
due to the absence of the real-space Berry curvature,
while it still poses nontrivial effects to Dirac electrons.

It is inadequate to treat the skyrmion texture on the TI surface perturbatively,
since it fails to incorporate the topological characteristics, namely the change in the Chern number inside the skyrmion.
We therefore need to solve the scattering problem of a Dirac electron by a single skyrmion non-perturbatively \cite{Ferreira,Ferreira_2014}.
The scattering process is characterized by the ``phase shifts'' of the eigenstates,
which are the angles quantifying how much the eigenstates are altered by the skyrmion \cite{R_G_Newton}.

\begin{figure}[tbp]
\includegraphics[width=6cm]{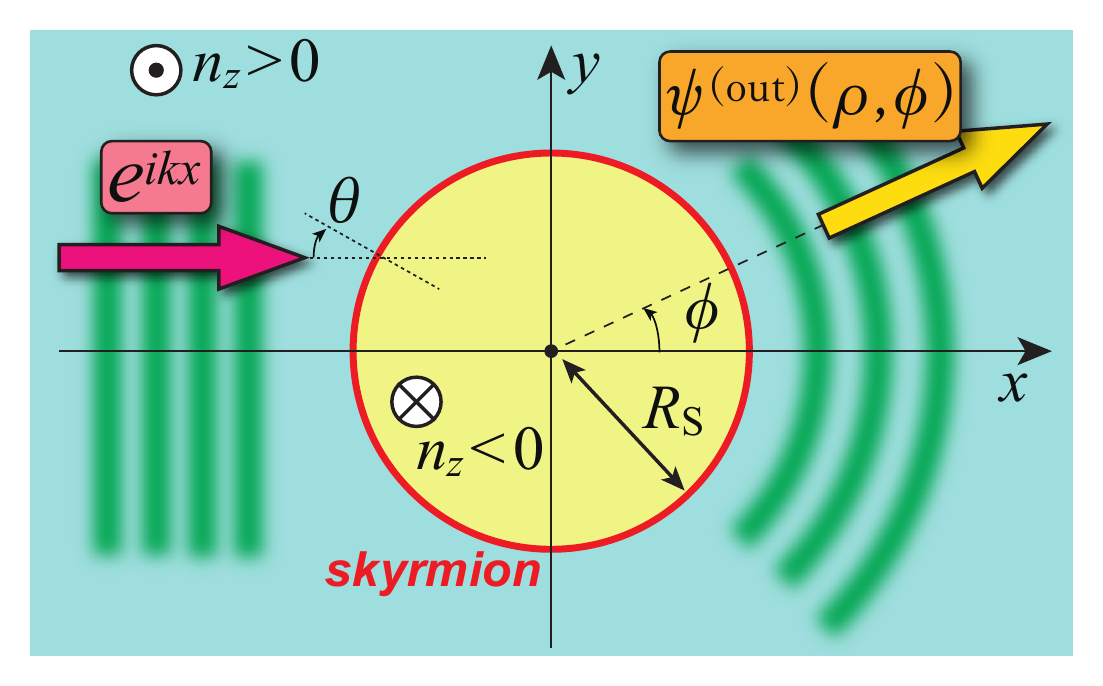} \\
\caption{
Schematic picture of the electron scattering by a single skyrmion.
An incident plane wave is scattered by a single skyrmion placed at the center,
leaving the outgoing spherical wave $\psi^\mathrm{(out)}(\rho,\phi)$.
$\theta$ denotes the incident angle to the skyrmion boundary.
}
\label{fig:configuration}
\end{figure}

Since the Hamiltonian is invariant under the simultaneous rotation in the real and spin spaces around the $z$-axis,
the total angular momentum $j = \pm 1/2, \pm 3/2, \cdots \in \mathbb{Z} +1/2$ serves as a good quantum number.
The eigenstate wave function with the angular momentum $j$ can be parametrized as
\begin{align}
\psi_j(\bfr) = 
\begin{pmatrix} u_j(\rho)e^{i (j-\frac{1}{2}) \phi} \\ v_j(\rho) e^{i (j+\frac{1}{2}) \phi} \end{pmatrix}.
\end{align}
The radial equations for $u_j(\rho)$ and $v_j(\rho)$ can be analytically solved under the hard-wall approximation
(see Appendix for detail).
The radial solution behaves as a cylindrical wave,
 with the asymptotic wave number $k_\mathrm{F} = v_\mathrm{F}^{-1}\sqrt{E_\mathrm{F}^2-\Delta^2}$ away from the skyrmion center $(\rho \rightarrow \infty)$,
for the conduction electron at the Fermi level $E_\mathrm{F}(>\Delta)$.
Using the phase shift $\delta_j$,
the asymptotic behavior can be expressed as
\begin{align}
\psi_{k_\mathrm{F},j}(\rho,\phi) \sim  \frac{1}{\sqrt{\rho}} \vecv{\sin\tfrac{\zeta}{2} \cos(\xi_j +\frac{\pi}{4}-\delta_j) e^{i(j-\frac{1}{2})\phi}}{-\cos\tfrac{\zeta}{2} \sin(\xi_j +\frac{\pi}{4}-\delta_j) e^{i(j+\frac{1}{2})\phi}}, \label{eq:asymptotic}
\end{align}
where $\xi_j \equiv k_\mathrm{F}\rho -\frac{\pi}{2}(j+\frac{1}{2})$
and $\cos\zeta \equiv d \equiv \Delta/E_\mathrm{F}$.
The skyrmion-free solution can be reproduced by setting $\delta_j =0$.
The phase shift $\delta_j$ is determined to make the solution continuous at the skyrmion edge $R_\mathrm{S}$,
given by
\begin{align}
\cot\delta_j = -\frac{1}{2d}\left[(1+d)T_{j+\frac{1}{2}}(\xi_\mathrm{F}) - (1-d)T_{j-\frac{1}{2}}(\xi_\mathrm{F}) \right] \label{eq:phase-shift}
\end{align}
for each angular momentum mode $j$,
with $\xi_\mathrm{F} \equiv k_\mathrm{F} R_\mathrm{S}$.
Here $T_l(\xi) \equiv Y_l(\xi) / J_l(\xi)$,
with $J_l$ the Bessel function and $Y_l$ the Neumann function.
As can be see from this relation,
the electron scattering behavior by the skyrmion is characterized by two dimensionless parameters,
$d \equiv \Delta/E_\mathrm{F}$ and $\xi_\mathrm{F} \equiv k_\mathrm{F} R_\mathrm{S}$.

The asymptotic behavior of the phase shift $\delta_j$ can be derived from that of the Bessel functions $J$ and $Y$,
the details of which are shown in the Appendix.
In the long-wavelength limit $\xi_\mathrm{F} \ll |j|$,
$\delta_j$ is given as
\begin{align}
\tan\delta_j = -\frac{2\pi d}{d+\sgn j} \frac{\tilde{j}}{(\tilde{j}!)^2} \left(\frac{\xi_\mathrm{F}}{2}\right)^{2\tilde{j}} + O \left(\xi_\mathrm{F}^{2\tilde{j}+2} \right), \label{eq:asymptotic-long-wavelength}
\end{align}
where $\tilde{j} \equiv |j|+1/2$.
Therefore, for fixed $\xi_\mathrm{F}$,
the scattering phase shift $\delta_j$ approaches zero as the angular momentum $|j|$ becomes larger than $\xi_\mathrm{F}$.
It can be well understood in the semiclassical picture:
if the impact parameter $|y| \sim |j|/k_\mathrm{F}$,
namely the distance from the scattering center to the incident trajectory,
is much larger than the skyrmion radius $R_\mathrm{S}$,
the incident particle is almost unaffected by the skyrmion.
In the short-wavelength limit $\xi_\mathrm{F} \gg |j|$, on the other hand,
$\delta_j$ can be approximated as
\begin{align}
\tan\delta_j = - \frac{d \cos(2\xi_\mathrm{F})}{(-1)^{j-\frac{1}{2}}+d \sin(2\xi_\mathrm{F})} + O(\xi_\mathrm{F}^{-1}), \label{eq:asymptotic-short-wavelength}
\end{align}
which implies that $\delta_j = \delta_{j+2}$ up to the difference of $O(\xi_\mathrm{F}^{-1})$.

If a plane wave with the incident wave vector $\boldsymbol{k}=(k_\mathrm{F},0)$ is scattered by the skyrmion placed at the origin,
as shown in Fig.~\ref{fig:configuration},
we can set up an ansatz for the total wave function away from the skyrmion $(\rho \rightarrow \infty)$ as
\begin{align}
\Psi(\boldsymbol{r}) \sim e^{ik_\mathrm{F} x}\vecv{-i\sin\frac{\zeta}{2}}{\cos\frac{\zeta}{2}} + \frac{e^{ik_\mathrm{F} \rho}}{\sqrt{\rho}} \vecv{f_\uparrow(\phi)}{f_\downarrow(\phi)}, \label{eq:wf-asymptotic}
\end{align}
where the first term comes from the incoming plane wave in the $x$-direction.
The second term corresponds to the outgoing scattered waves,
which we denote $\psi^\mathrm{(out)}(\rho,\phi)$ here.
Since the total wave function on the right hand side can be given as a linear combination of the eigenfunctions $\{ \psi_{k_\mathrm{F},j} \}$,
we can estimate the scattering amplitudes $f_{\uparrow/\downarrow}(\phi)$
by the partial wave decomposition:
by decomposing both sides of Eq.~(\ref{eq:wf-asymptotic})
and comparing the radial parts for each partial wave component,
we obtain the scattering amplitudes
\begin{align}
\vecv{f_\uparrow(\phi)}{f_\downarrow(\phi)} = \frac{1}{\sqrt{2\pi k_\mathrm{F}}}e^{i\frac{\pi}{4}} \Phi(\phi) \vecv{\sin\frac{\zeta}{2} e^{-i\phi/2}}{-\cos\frac{\zeta}{2} e^{i\phi/2}}, \label{eq:scattering-amplitude-f}
\end{align}
where the angular function $\Phi(\phi)$ is composed of the phase shifts as
\begin{align}
\Phi(\phi) = \sum_{j} e^{ij\phi}(1-e^{-2i\delta_j}),
\end{align}
with $j \in \mathbb{Z} +1/2$.
As shown in Eq.~(\ref{eq:asymptotic-long-wavelength}),
since the phase shift approaches zero for sufficiently large $j$,
namely $|j| \gg k_\mathrm{F} R_\mathrm{S}$,
we shall limit the partial waves to $|j| < 2\xi_\mathrm{F} +1$ in the calculations below.

\begin{figure}[tbp]
\includegraphics[width=7.2cm]{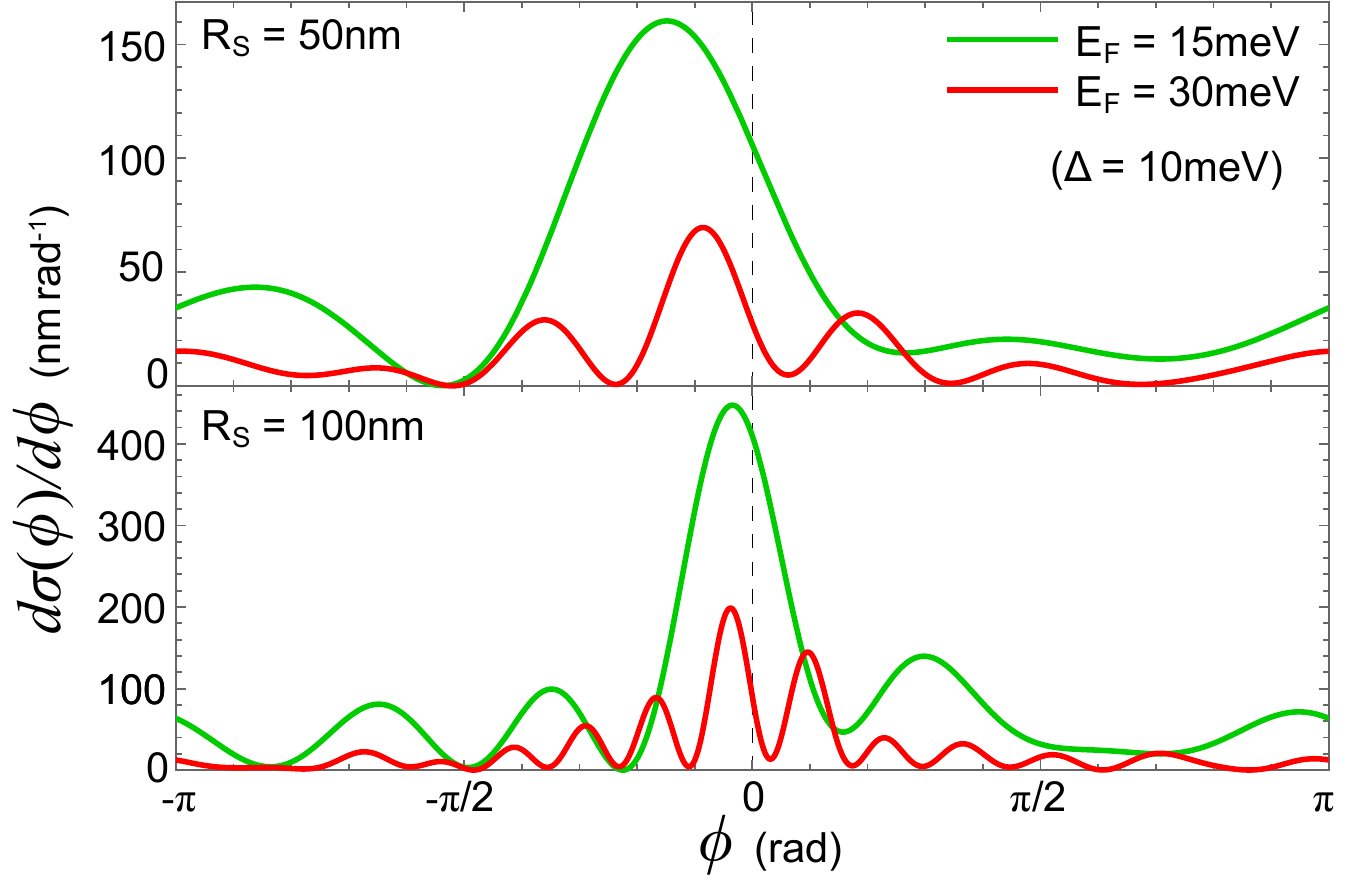} \\
\caption{The angular profile of the differential cross section $d\sigma(\phi)/d\phi$ induced by a hard-wall skyrmion,
with its radius $R_\mathrm{S}=50\mathrm{nm}$ in the upper panel and $R_\mathrm{S}=100\mathrm{nm}$ in the lower panel.
}
\label{fig:cross_section}
\end{figure}

Using the outgoing wave function $\psi^\mathrm{(out)}$ obtained above,
 the differential cross section,
namely the ratio between the incoming current density $j^\mathrm{(in)}$
 and the outgoing flux $j^\mathrm{(out)}(\phi) \rho d\phi$ toward the angle $\phi$,
is given by
\begin{align}
\frac{d\sigma(\phi)}{d\phi} = \frac{j^\mathrm{(out)}\rho}{j^\mathrm{(in)}} = \frac{2E_\mathrm{F}}{v_\mathrm{F} k_\mathrm{F}} \im\left[e^{-i\phi} f_\uparrow^* f_\downarrow \right] = \frac{|\Phi(\phi)|^2}{2\pi k_\mathrm{F}} .
\end{align}
The angular profile of the differential cross section $d\sigma(\phi)/d\phi$ is shown in Fig.~\ref{fig:cross_section},
for several Fermi energies $E_\mathrm{F}$ and skyrmion radii $R_\mathrm{S}$.
In contrast to normal scattering processes,
such as the scattering of Schr\"{o}dinger electrons by a symmetric Coulomb potential,
the scattering process shows a skewness, asymmetric about $\phi=0$.

\modify{The strengths of the whole scattering, the back scattering, and the skew scattering,
are measured by the integrals}
\begin{align}
\begin{pmatrix} F \\ F_\parallel \\ F_\perp \end{pmatrix} &\equiv \int \frac{d\phi}{2\pi} \ |\Phi(\phi)|^2 \begin{pmatrix} 1 \\ 1-\cos\phi \\ \sin\phi \end{pmatrix} = \sum_j \begin{pmatrix} 2\sin^2\delta_j \\ 2\sin^2 \bar{\delta}_j \\ \sin 2\bar{\delta}_j \end{pmatrix},
\end{align}
respectively,
with $\bar{\delta}_j = \delta_{j+1} - \delta_j$.
It is obvious that the back and skew scattering strengths do not exceed the whole scattering strength,
i.e. $F_\parallel,|F_\perp| < F$.
If the skyrmion size is smaller than the electron wavelength ($\xi_\mathrm{F} \ll 1$),
the asymptotic behavior for $\delta_j$ in Eq.~(\ref{eq:asymptotic-long-wavelength}) applies for all the partial wave modes.
The lowest angular momentum modes $(j=\pm 1/2)$ give the dominant contribution to $F$,
given by
\begin{align}
F \simeq 2 \sum_\pm \sin^2 \delta_{\pm 1/2} = \frac{16\pi^2 d^2}{1-d^2}\xi_\mathrm{F}^4 + O(\xi_\mathrm{F}^6).
\end{align}
On the other hand, if the skyrmion is large enough $(\xi_\mathrm{F} \gg 1)$,
the modes with $|j| \ll \xi_\mathrm{F}$ yield the dominant contribution to $F$.
Using the asymptotic behavior in Eq.~(\ref{eq:asymptotic-short-wavelength}),
the leading-order contribution to the total scattering amplitude is given as
\begin{align}
F \simeq 2 \sum_{|j|<\xi_\mathrm{F}} \sin^2 \delta_j = \frac{4\xi_\mathrm{F} d^2(1+d^2)\cos^2(2\xi_\mathrm{F})}{(1+d^2)^2-4d^2\sin^2(2\xi_\mathrm{F})} + O(\xi_\mathrm{F}^0).
\end{align}
If the exchange gap $\Delta$ is small enough compared with the electron energy $E_\mathrm{F}$,
i.e. $d \ll 1$,
it further reduces to the simplified behavior,
$F = 4 \xi_\mathrm{F} d^2 \cos^2(2\xi_\mathrm{F}) + O(\xi_F^0,d^4)$.
Both the backscattering strength $F_\parallel$ and the skew scattering strength $|F_\perp|$
are limited by this upper bound,
yielding
\begin{align}
F_\parallel,|F_\perp|  \lesssim 4 \xi_\mathrm{F} d^2 + O(\xi_F^0,d^4).
\end{align}
This asymptotic relation shall be used for the qualitative estimation
of the transport properties in the next section.

The scattering skewness shown here can be traced back to the geometric phase acquired at the skyrmion boundary,
which is {characteristic to Dirac electrons}.
When an electron is transmitted through the boundary between two regions with opposite magnetizations,
{the scattered wave acquires a phase factor $e^{i\gamma(\theta)}$ relative to the incident wave,}
depending on its incident angle $\theta$.
In the electron scattering process treated here,
the incident angle $\theta$ at the skyrmion boundary depends on the impact parameter $y$,
{at sufficiently long wavelength} such that the uncertainty in $y$ can be neglected (see Fig.~\ref{fig:configuration}).
Therefore, the phase factor $e^{i\gamma(\theta)}$ depends on $y$;
{since this phase can be expanded as $\gamma(\theta(y)) = \pi + \tilde{k}y + O(y^2)$ around the scattering center,
the scattered wave bears the transverse component $e^{i \tilde{k} y}$,}
leading to the scattering skewness
(see Appendix for detail).

\section{Semiclassical transport analysis}
The electron scattering by skyrmions alters the electronic transport properties,
the conductivity in particular.
Here we set up a uniformly magnetized 2D system with an ensemble of skyrmions, with the number density $n_\mathrm{S}$.
As long as the skyrmion distribution is random and dilute enough,
quantum interference from multiple scattering processes is ruled out.
\modify{Here the interplay between skyrmion scattering and other scattering process,
such as skew or side-jump scattering at normal impurities in the presence of SOC,
is quite small,
so that scattering by skyrmions contributes to the conductivity additively.}
Hence we estimate this additive contribution by the semiclassical Boltzmann theory.

The electron transport driven by an electric field $\boldsymbol{E} = E_x \boldsymbol{e}_x$ is described by the Boltzmann equation,
\begin{align}
-e \boldsymbol{E} \cdot \nabla_{\bfk}f(\bfk) = \left(\frac{df(\bfk)}{dt}\right)_\mathrm{coll}, \label{eq:Boltzmann}\end{align}
for the steady-state electron distribution $f(\boldsymbol{k})$.
The scattering process is {incorporated in} the collision term
\begin{align}
& \left(\frac{df(\bfk)}{dt}\right)_\mathrm{coll} = -\frac{1}{\tau_0} \left[ f(\bfk) -f_0(\bfk) \right] \\
& \quad \quad \quad \quad -n_\mathrm{S} \int d\phi |\boldsymbol{v}(\bfk)| \left[ \frac{d\sigma(\phi)}{d\phi}f(\bfk) - \frac{d\sigma(-\phi)}{d\phi}f(\bfk') \right], \nonumber
\end{align}
with the velocity $\boldsymbol{v}(\bfk) = \boldsymbol{\nabla}_{\bfk} E(\bfk)$,
where we omit the intrinsic contribution from the $\bfk$-space Berry curvature.
The second term describes the skyrmion-induced scattering process from momentum $\bfk$ into $\bfk'$ and vice versa,
with $\phi$ the angle between them.
We also introduce the backscattering effect by normal impurities phenomenologically by the first term,
in terms of relaxation-time approximation,
with the transport relaxation time $\tau_0$ and the equilibrium distribution $f_0(\bfk)$.
\modify{The effect of skew and side-jump scattering processes by normal impurities are so far neglected,
whose contributions to the conductivity shall be incorporated in our discussion later.}
The momenta $\bfk$ and $\bfk'$ are limited to the Fermi surface, i.e. $|\bfk| = |\bfk'|=k_\mathrm{F}$,
since the scattering here is elastic.

The steady-state distribution $f(\bfk)$ can be obtained
up to the linear response in the electric field $\boldsymbol{E}$.
Taking the ansatz \cite{Ferreira_2014}
\begin{align}
\delta f(\bfk) &= \frac{\partial f_0}{\partial E}\biggl|_{E(\bfk)} \left[ (\bfk\cdot e\boldsymbol{E})C_k + (\boldsymbol{e}_z \times \bfk \cdot e\boldsymbol{E})D_k \right]
\end{align}
for the deviation of the distribution function $\delta f(\bfk) = f(\bfk) - f_0(\bfk)$,
we can straightforwardly solve the Botlzmann equation at zero temperature.
Since we are neglecting the interplay between skyrmion and other scattering processes,
we evaluate the solution up to the first order in the skyrmion concentration $n_\mathrm{S}$
(we shall check the validity of this approximation in the numerical calculation later),
yielding
\begin{align}
\begin{pmatrix} C_k \\ D_k \end{pmatrix}
= \frac{v_\mathrm{F}^2}{E_\mathrm{F}}
\begin{pmatrix} \tau_0 - \tau_0^2 \tau_\parallel^{-1} \\ -\tau_0^2 \tau_\perp^{-1} \end{pmatrix}.
\end{align}
Here $\tau_\parallel$ and $\tau_\perp$ are the ``scattering times''
characterizing the backscattering and skew scattering effects
from the randomly distributed skyrmions,
respectively, given by
\begin{align}
\tau_X^{-1} = n_\mathrm{S} \frac{v_\mathrm{F}^2}{E_\mathrm{F}} F_X. \quad (X=\parallel,\perp)
\end{align}

\begin{figure}[tbp]
\begin{tabular}{c}
\includegraphics[width=7cm]{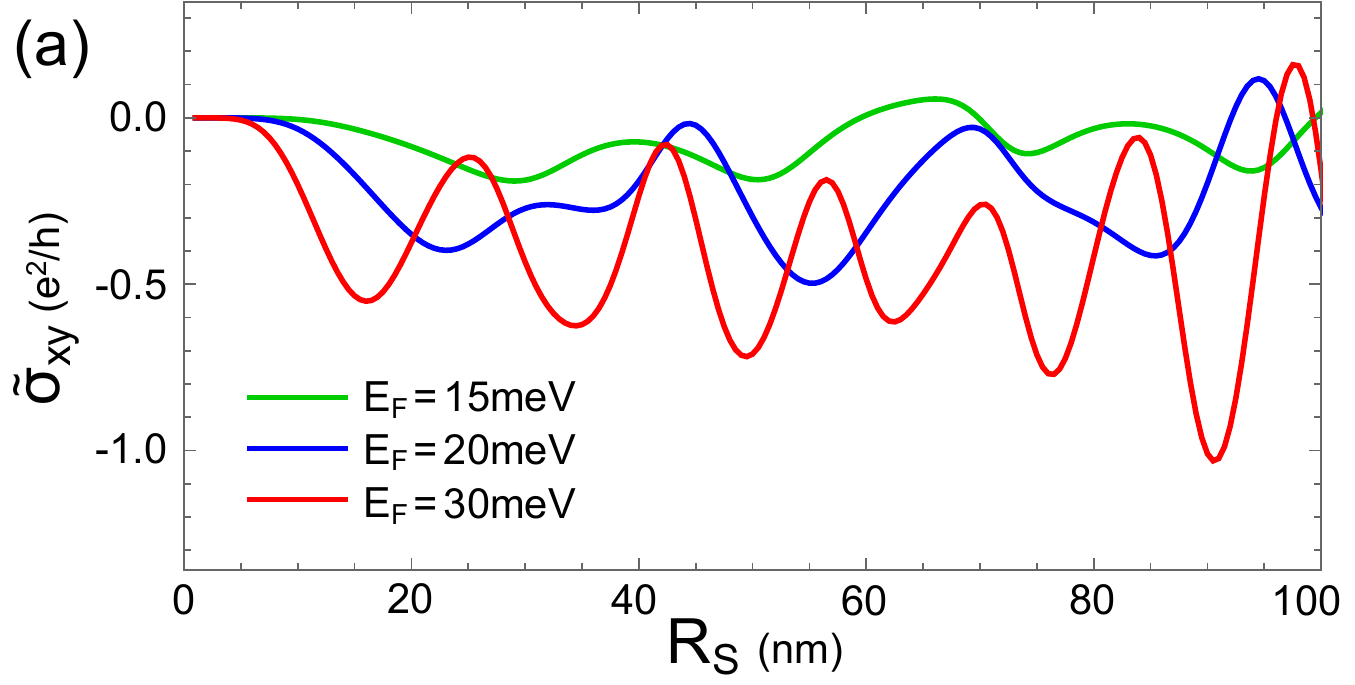} \\
\includegraphics[width=6.8cm]{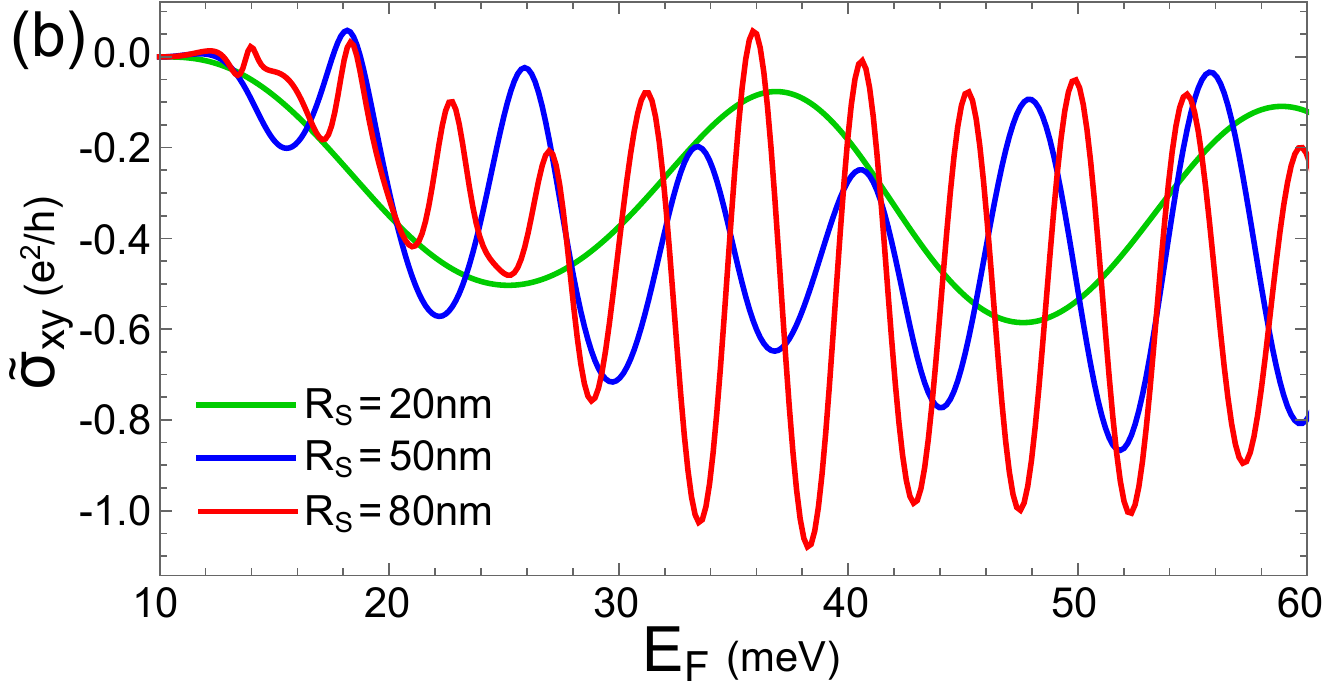}
\end{tabular}
\caption{The behavior of the skyrmion contribution to the Hall conductivity, $\tilde{\sigma}_{xy}$,
(a) with the Fermi energy $E_\mathrm{F}$ fixed and the skyrmion radius $R_\mathrm{S}$ varied,
and (b) with $R_\mathrm{S}$ fixed and $E_\mathrm{F}$ varied,
under the magnetic exchange energy $\Delta = 10 \mathrm{meV}$.
}
\label{fig:conductivity}
\end{figure}

Once we obtain the distribution shift $\delta f(\bfk)$ driven by the electric field,
we can estimate its extrinsic contribution to the current by
$\tilde{\boldsymbol{j}} = -(e/V) \sum_{\bfk} \boldsymbol{v}(\bfk) \delta f(\bfk)$,
leading to the conductivity coefficients
\begin{align}
\begin{pmatrix} \tilde{\sigma}_{xx} \\ \tilde{\sigma}_{xy} \end{pmatrix}
= \frac{e^2}{4\pi} \frac{E_\mathrm{F}^2 -\Delta^2}{E_\mathrm{F}}
\begin{pmatrix} \tau_0 - \tau_0^2 \tau_\parallel^{-1} \\ \tau_0^2 \tau_\perp^{-1} \end{pmatrix}
,
\end{align}
up to the first order in $n_\mathrm{S}$.
\modify{The longitudinal part $\tilde{\sigma}_{xx}$
consists of the backscattering effect by the normal scatterers, proportional to $\tau_0$,
and that by the skyrmions, proportional to $\tau_0^2 \tau_\perp^{-1}$.
}
On the other hand, \modify{as we have mentioned},
$\tilde{\sigma}_{xy}$ obtained here is the additive contribution to the Hall conductivity induced by the skyrmions;
other contributions, such as the ordinary Hall effect contribution from the external magnetic field,
the intrinsic contribution from the momentum-space Berry curvature \cite{Qi_Wu_Zhang},
and the extrinsic (skew and side-jump) contribution from the normal scatterers \cite{Sinitsyn,Sinitsyn_review,Culcer,Lu_2013,Titov},
come additively to the Hall conductivity as well,
which shall be discussed at the end of this section.

Now we calculate the skyrmion contribution to the Hall conductivity $\tilde{\sigma}_{xy}$
as functions of Fermi energy $E_\mathrm{F}$ and the skyrmion radius $R_\mathrm{S}$, which are shown in Fig.~\ref{fig:conductivity}.
We take screened Coulomb impurities as the normal scatterers,
which give the transport relaxation time $\tau_0$ proportional to
 $n_\mathrm{c}^{-1/2}$,
with $n_\mathrm{c} = k_\mathrm{F}^2/4\pi$ the carrier density \cite{Nomura_MacDonald_2006,Nomura_MacDonald_2007,Ando_JPSJ,Hwang_DasSarma}.
Here we employ $\tau_0$ shown in Ref.~\onlinecite{Hwang_DasSarma},
which was calculated for graphene deposited on $\mathrm{SiO}_2$ substrate,
yielding $\tau_0 = 0.6 \mathrm{ps} \times (k_\mathrm{F}/0.1\mathrm{nm}^{-1})$ under the impurity density $n_\mathrm{i} = 10^{11} \mathrm{cm}^{-2}$.
The skyrmion number density $n_\mathrm{S}$ is fixed to $2.89\times 10^9\mathrm{cm}^{-2}$,
corresponding to the triangular skyrmion lattice with the spacing $200\mathrm{nm}$.
\modify{With those parameters, we can estimate the skyrmion scattering time $\tau_X \ (X=\parallel,\perp)$
around $E_\mathrm{F} \sim 50\mathrm{meV}$ as}
\begin{align}
\tau_X = \frac{E_\mathrm{F}}{n_\mathrm{S} v_\mathrm{F}^2 F_X} \sim \frac{10 \mathrm{ps}}{F_X}.
\end{align}
\modify{At the skyrmion size $R_\mathrm{S}$ around $100\mathrm{nm}$,
$F_X$ can be roughly estimated as $F_X \lesssim 4 \xi_\mathrm{F} d^2 \sim 2$,
leading to $\tau_X \gtrsim 5\mathrm{ps}$.
Therefore, the skyrmion scattering effect $\tau_X^{-1}$ is weaker than the normal impurity effect $\tau_0^{-1}$,
so that the linear approximation by $n_\mathrm{S}$ (or $\tau_X^{-1}$) will be reliable in this calculation.}

Our calculation results show that the contribution to the Hall conductivity is suppressed at small $E_\mathrm{F}$ around the band bottom.
\modify{Such a behavior can be traced back to two reasons:
since spin-momentum locking becomes less significant around the band bottom,
the phase factor effect at the skyrmion boundary becomes weaker, suppressing the skew scattering effect.
Moreover, the Coulomb impurities give a larger scattering rate at lower electron concentration,
which makes the skyrmion effect relatively small.
On the other hand, if the chemical potential is set far away from the bandgap,
the band structure asymptotically reaches the gapless Dirac dispersion
so that the electron eigenstates will become insensitive to the magnetic texture,
suppressing the skyrmion effect.
Thus we need a moderate chemical potential to obtain a sizable skyrmion contribution to the Hall conductivity,
which is around $50\mathrm{meV}$ in our calculation.}

\modify{We also find a quantum oscillation-like behavior in $\tilde{\sigma}_{xy}$ under the modulation of
the skyrmion size $R_\mathrm{S}$ and the Fermi energy $E_\mathrm{F}$.
We expect that this oscillation may originate from the electron resonance states formed inside the skyrmion,
in which an electron standing wave should be formed within the skyrmion diameter.
Such resonance states become less stable under a smooth magnetic texture without the hard-wall approximation,
which may relax the oscillation behavior.}

From the results of Ref.~\onlinecite{Sinitsyn},
we can make an order estimation of the normal impurity contribution to the Hall conductivity
to compare it with the skyrmion contribution,
which is beyond the scope of our Boltzmann analysis.
If the impurity distribution is Gaussian,
its effect on the Hall conductivity is insensitive to the impurity potential strength
and scales by $\lesssim (e^2/4\pi)(\Delta/E_\mathrm{F})$ for $E_\mathrm{F} \gg \Delta$.
If the distribution is beyond Gaussian,
it acquires an additional term from the third-order moment of the impurity distribution,
corresponding to the skew scattering effect by the impurities.
It is known to be inversely proportional to the scattering strength,
scaling as $\sim (e^2/4\pi)(\Delta/n_\mathrm{i} V_\mathrm{i})$,
where $V_\mathrm{i}$ parametrizes the impurity potential in terms of delta-function potential.
At high $E_\mathrm{F}$, the impurity potential $V_\mathrm{i}$ gets reduced by strong screening,
leading to enhancement of this third-order moment contribution.
For $E_\mathrm{F} \sim 50 \mathrm{meV}$ and $n_\mathrm{i} \sim 10^{11}\mathrm{cm}^{-2}$,
the denominator $n_\mathrm{i} V_\mathrm{i}$ is around $10 \mathrm{meV}$,
estimated from the screened Coulomb potential \cite{Hwang_DasSarma}.
Therefore, the skyrmion contribution to the anomalous Hall conductivity
is comparable or even superior to the normal impurity contribution at $E_\mathrm{F} \sim 50 \mathrm{meV}$,
which is possibly captured in transport measurements.


\section{Discussion}
In this paper, we have investigated the electronic transport properties on TI surfaces in the presence of magnetic skyrmions.
We have demonstrated that the skyrmions give rise to the extrinsic AHE,
originating from the interplay between the magnetic texture and the Dirac band structure.
Variation of the Fermi energy $E_\mathrm{F}$ (or the Fermi momentum $k_\mathrm{F}$) can be experimentally realized
by tuning the gate voltage {or the carrier doping},
while the skyrmion size can also be modulated by external magnetic fields.

It should be noted that, in realistic magnetic materials,
an external magnetic field is essential to stabilize skyrmions.
In $\mathrm{Cu}_2 \mathrm{O} \mathrm{Se} \mathrm{O}_3$, for instance,
a magnetic field of $\sim 0.05\mathrm{T}$ is required,
which corresponds to the magnetic length $\sim 100\mathrm{nm}$.
The skyrmion size should not exceed this scale;
otherwise the Landau quantization inside the skyrmion takes place and the scattering property obtained in this paper
is no longer reliable.

We have so far relied on the hard-wall approximation on the vortex-type skyrmion texture.
Such an approximation is justified under strong out-of-plane magnetic anisotropy,
which stabilizes out-of-plane magnetization pattern.
On the other hand, our results do not apply to the hedgehog skyrmion,
since it cannot be gauged away due to the emergent magnetic flux.
Such type of skyrmion is expected to be induced by frustrated exchange interactions or four-spin exchange interactions \cite{Nagaosa_Tokura},
whose effect on the electron transport on TI surfaces remains an open question.

\acknowledgments{
The authors acknowledge T.~Chiba, C.~Lee, Y.~Ominato, and K.~Sato for fruitful discussions.
Y.~A. is supported by JSPS KAKENHI Grant Number JP17K14316.
K.~N. is supported by JSPS KAKENHI Grant Numbers JP15H05854 and JP17K05485.}

\appendix

\section{Eigenstate wave functions under a single skyrmion}
In this part, we show the detailed derivation of the eigenstate wave functions under a single skyrmion,
which leads to the phase shift [Eq.~(3)].
For the partial wave component with the total angular momentum $j (= \pm 1/2, \pm 3/2, \cdots)$,
we can set up the ansatz
\begin{align}
\psi_j(\rho,\phi) = \vecv{u_j(\rho) e^{i l_\uparrow\phi}}{v_j(\rho) e^{i l_\downarrow\phi}},
\end{align}
with $l_{\uparrow/\downarrow} \equiv j\mp 1/2$ the orbital angular momentum for each component.
With this ansatz, the Dirac equation $\mathcal{H} \psi = E\psi$ reduces to the radial differential equation,
\begin{align}
\matr{-\Delta n_z(\rho) -E}{v_\mathrm{F}(-\partial_\rho - l_\downarrow \rho^{-1})}{v_\mathrm{F}(\partial_\rho - l_\downarrow \rho^{-1})}{\Delta n_z(\rho) -E} \vecv{u_j(\rho)}{v_j(\rho)} =0,
\end{align}
under the hard-wall approximation on the skyrmion texture.

Let us solve the equation without skyrmion, namely with $n_z(\rho)=1$ in the entire system, as the starting point.
Here the equation reads
\begin{align}
\matr{-\Delta-E}{v_\mathrm{F}(-\partial_\rho - l_\downarrow \rho^{-1})}{v_\mathrm{F}(\partial_\rho - l_\downarrow \rho^{-1})}{\Delta -E} \vecv{u_j(\rho)}{v_j(\rho)} =0. \label{eq:radial-equation}
\end{align}
Using the relation
\begin{align}
v_j(\rho) &= -\frac{v_\mathrm{F}}{\Delta-E}\left(\partial_\rho -\frac{l_{\uparrow}}{\rho}\right) u_{j}(\rho), \label{eq:u-v}
\end{align}
we obtain the second-order differential equation for $u$,
\begin{align}
\left[\partial_\rho^2 +\frac{1}{\rho}\partial_\rho +\frac{E^2-\Delta^2}{v_\mathrm{F}^2} -\frac{l_\uparrow^2}{\rho^2} \right] u_{j}(\rho) =0. \label{eq:diff-uniform}
\end{align}
With the radial wavenumber $k \equiv v_\mathrm{F}^{-1} \sqrt{E^2 - \Delta^2}$
and the change of variable $\xi \equiv k \rho$,
Eq.~(\ref{eq:diff-uniform}) reduces to the well-known Bessel's differential equation,
\begin{align}
\left[\partial_\xi^2 - \frac{1}{\xi}\partial_\xi +1 - \frac{l_\uparrow^2}{\xi^2} \right] u_{j}(\xi) =0,
\end{align}
whose linearly independent solutions are given by the Bessel function $J_{l_\uparrow}(\xi)$ and the Neumann function $Y_{l_\uparrow}(\xi)$.
Since $Y_{l_\uparrow}$ is ruled out for solution due to the singularity at $\rho =0$,
we can set $u_{j}(\rho) = C_{k,j}^{(+)} J_{l_\uparrow}(k\rho)$ with $C$ the normalization constant.
Substituting this solution to Eq.~(\ref{eq:u-v}), we obtain the eigenstate wave function
\begin{align}
\chi_{k,j}^{(+)}(\rho,\phi) &= C_{k,j}^{(+)} \vecv{J_{l_\uparrow}(k\rho) e^{i l_\uparrow\phi}}{-\frac{v_\mathrm{F} k}{E-\Delta}J_{l_\downarrow}(k\rho) e^{i l_\downarrow \phi}},
\end{align}
where we have used the recurrence relation
\begin{align}
\left(\partial_\xi \mp \frac{l}{\xi} \right) J_l(\xi) = \mp J_{l \pm 1}(\xi).
\end{align}
The normalization condition
\begin{align}
\int_0^{\infty} d\rho \ \rho \int_0^{2\pi} d\phi \left[\chi_{k,j}^{(+)}(\bfr)\right]^\dag \chi_{k',j}^{(+)}(\bfr) =\delta(k-k')
\end{align}
fixes the normalization constant as
\begin{align}
2\pi \left|C_{k,j}^{(+)}\right|^2 \frac{1}{k} \left[1+\left(\frac{v_\mathrm{F} k}{E-\Delta}\right)^2\right] =1,
\end{align}
where we have used the orthogonality relation
\begin{align}
\int_0^{\infty} d\rho \ \rho J_l(k\rho) J_l(k'\rho) = \frac{1}{k}\delta(k-k').
\end{align}
Therefore, the normalized eigenstate wave function is given by
\begin{align}
\chi_{k,j}^{(+)}(\rho,\phi) &= \sqrt{\frac{k}{2\pi}} \vecv{\sin\tfrac{\zeta}{2} J_{l_\uparrow}(k\rho) e^{i l_\uparrow\phi}}{-\cos\tfrac{\zeta}{2} J_{l_\downarrow}(k\rho) e^{i l_\downarrow \phi}}. \label{eq:free-solution}
\end{align}
Here the angle $\zeta$ is defined by $E = \Delta/\cos \zeta$,
so that it should satisfy
\begin{align}
\frac{v_\mathrm{F} k}{E-\Delta} = \frac{\tan\zeta}{\sec\zeta -1} = \frac{\sin\zeta}{1-\cos\zeta} =\cot\frac{\zeta}{2}.
\end{align}
We should note here the relation
\begin{align}
J_{-l}(\xi) = (-1)^l J_l(\xi)
\end{align}
for integer $l$.
If the magnetization is flipped, i.e. $n_z(\rho) =-1$,
we should substitute $\zeta$ by $\pi-\zeta$,
leading to the eigenfunction
\begin{align}
\chi_{k,j}^{(-)}(\rho,\phi) &= \sqrt{\frac{k}{2\pi}} \vecv{\cos\tfrac{\zeta}{2} J_{l_\uparrow}(k\rho) e^{i l_\uparrow\phi}}{-\sin\tfrac{\zeta}{2} J_{l_\downarrow}(k\rho) e^{i l_\downarrow \phi}}. \label{eq:free-solution-negative}
\end{align}


In the presence of a hard-wall skyrmion, i.e.
\begin{align}
n_z(\rho<R_\mathrm{S}) = -1 ; \quad n_z(\rho>R_\mathrm{S}) = +1,
\end{align}
we should solve Eq.~(\ref{eq:radial-equation}) inside and outside the skyrmion separately,
and connect the solution at the boundary.
Inside the skyrmion, the solution is given by Eq.~(\ref{eq:free-solution-negative}).
Outside the skyrmion, on the other hand, the solutions is not simply given by Eq.~(\ref{eq:free-solution}),
since we do not need to respect the normalizability at the origin.
The Neumann counterpart of $\chi_{k,j}^{(+)}$, given by
\begin{align}
\eta_{k,j}^{(+)}(\rho,\phi) &= \sqrt{\frac{k}{2\pi}} \vecv{\sin\tfrac{\zeta}{2} Y_{l_\uparrow}(k\rho) e^{i l_\uparrow\phi}}{-\cos\tfrac{\zeta}{2} Y_{l_\downarrow}(k\rho) e^{i l_\downarrow \phi}}, \label{eq:free-solution-neumann}
\end{align}
also contributes to the solution.
Thus the solution outside the skyrmion should be given by the linear combination,
\begin{align}
\psi_{k,j}^{(+)}(\rho,\phi) = A_{k,j} \chi_{k,j}^{(+)}(\rho,\phi) + B_{k,j} \eta_{k,j}^{(+)}(\rho,\phi),
\end{align}
where the coefficients $A_{k,j}$ and $B_{k,j}$ are determined by the boundary condition
\begin{align}
\chi_{k,j}^{(-)}(\rho\rightarrow R_\mathrm{S},\phi) = \psi_{k,j}^{(+)}(\rho \rightarrow R_\mathrm{S},\phi).
\end{align}
Thus we obtain the coupled equations for those coefficients,
\begin{align}
\matr{J_{l_\uparrow}(kR_\mathrm{S})}{J_{l_\downarrow}(kR_\mathrm{S})}{Y_{l_\uparrow}(kR_\mathrm{S})}{Y_{l_\downarrow}(kR_\mathrm{S})}
\vecv{A_{k,j}}{B_{k,j}}
=
\vecv{\cot\tfrac{\zeta}{2} J_{l_\uparrow}(kR_\mathrm{S})}{ \tan\tfrac{\zeta}{2} J_{l_\downarrow}(kR_\mathrm{S})}.
\end{align}
The equations can be exactly solved, yielding
\begin{align}
A_{k,j} &= \frac{\cot\tfrac{\zeta}{2}T_{l_\downarrow}(k R_\mathrm{S}) - \tan\tfrac{\zeta}{2}T_{l_\uparrow}(k R_\mathrm{S})}{T_{l_\downarrow}(k R_\mathrm{S}) - T_{l_\uparrow}(k R_\mathrm{S})} \\
B_{k,j} &= \frac{\tan\tfrac{\zeta}{2} - \cot\tfrac{\zeta}{2}}{T_{l_\downarrow}(k R_\mathrm{S}) - T_{l_\uparrow}(k R_\mathrm{S})},
\end{align}
with
\begin{align}
T_{l}(\xi) \equiv Y_l(\xi) / J_l(\xi).
\end{align}
In order to meet the orthogonality relation
\begin{align}
\int_0^{\infty} d\rho \ \rho \int_0^{2\pi} d\phi \left[\psi_{k,j}(\bfr)\right]^\dag \psi_{k',j}(\bfr) =\delta(k-k'),
\end{align}
the whole wave function $\psi_{k,j}(\bfr)$ should be divided by the factor $C_{k,j} \equiv \sqrt{A_{k,j}^2+B_{k,j}^2}$,
yielding the final solution
\begin{align}
\psi_{k,j}(\rho,\phi) =
\begin{cases}
 \frac{1}{C_{k,j}} \chi_{k,j}^{(-)}(\rho,\phi) & (\rho<R_\mathrm{S}) \\
 \frac{A_{k,j}}{C_{k,j}} \chi_{k,j}^{(+)}(\rho,\phi) +  \frac{B_{k,j}}{C_{k,j}} \eta_{k,j}^{(+)}(\rho,\phi) & (\rho > R_\mathrm{S})
\end{cases} .
\end{align}
The total wave function $\Psi(\bfr)$ is given by the linear combination of the eigenstates,
namely
\begin{align}
\Psi(\bfr) = \sum_{j \in \mathbb{Z}+1/2} \alpha_j \psi_{k,j}(\rho,\phi).
\end{align}

The asymptotic behavior of this solution in the limit $\rho \rightarrow \infty$ is important in deriving the scattering amplitude.
Here we fix the incident wave number to the Fermi momentum $k_\mathrm{F} = v_\mathrm{F}^{-1} \sqrt{E_\mathrm{F}^2-\Delta^2}$.
Using the asymptotic behavior of the Bessel functions for $\xi/|l| \gg 1$,
\begin{align}
J_l(\xi) &\sim \sqrt{\frac{2}{\pi\xi}} \left[ \cos \left(\xi -\frac{l\pi}{2} -\frac{\pi}{4}\right) +O(\xi^{-1}) \right] \label{eq:asymptotic-J} \\
Y_l(\xi) &\sim \sqrt{\frac{2}{\pi\xi}} \left[ \sin \left(\xi -\frac{l\pi}{2} -\frac{\pi}{4}\right) +O(\xi^{-1}) \right],\label{eq:asymptotic-Y}
\end{align}
their linear combination satisfies
\begin{align}
& J_l(\xi) \cos\delta + Y_l(\xi) \sin\delta \nonumber \\
& \sim \sqrt{\frac{2}{\pi\xi}} \left[ \cos \left(\xi -\frac{l\pi}{2} -\frac{\pi}{4} -\delta \right) +O(\xi^{-1}) \right],
\end{align}
where $\delta$ serves as the phase shift of the cylindrical wave in comparison with the original wave $J_l$.
Thus, by introducing the phase shift
\begin{align}
\delta_j & \equiv \arctan \frac{B_{k_\mathrm{F},j}}{A_{k_\mathrm{F},j}} \\
 & = -\arctan \frac{\cos^2\tfrac{\zeta}{2} - \sin^2\tfrac{\zeta}{2}}{\cos^2\tfrac{\zeta}{2}T_{l_\downarrow}(\xi_\mathrm{F}) - \sin^2\tfrac{\zeta}{2}T_{l_\uparrow}(\xi_\mathrm{F})} \\
 & = -\arctan \frac{2\cos\zeta}{(1+\cos\zeta)T_{l_\downarrow}(\xi_\mathrm{F}) - (1-\cos\zeta)T_{l_\uparrow}(\xi_\mathrm{F})} \\
 & = -\arctan \frac{2d}{(1+d)T_{l_\downarrow}(\xi_\mathrm{F}) - (1-d)T_{l_\uparrow}(\xi_\mathrm{F})},
\end{align}
with $\xi_\mathrm{F} = k_\mathrm{F} R_\mathrm{S}$ and $d = \Delta/E_\mathrm{F}$,
we obtain the asymptotic behavior
\begin{align}
& \psi_{k_\mathrm{F},j}(\rho \rightarrow \infty, \phi) \label{eq:Psi-asymptotic} \\ 
 &  \sim \sqrt{\frac{2}{\pi k_\mathrm{F} \rho}} \vecv{\sin\tfrac{\zeta}{2}\cos\left[k_\mathrm{F} \rho -\tfrac{\pi}{2}l_\uparrow -\tfrac{\pi}{4} -\delta_{j}\right] e^{i l_\downarrow \phi}}{-\cos\tfrac{\zeta}{2}\cos\left[k_\mathrm{F} \rho -\tfrac{\pi}{2}l_\downarrow -\tfrac{\pi}{4}-\delta_{j}\right] e^{i l_\uparrow \phi}}. \nonumber
\end{align}

For large angular momentum $(|j| \gg k_\mathrm{F} R_\mathrm{S} = \xi_\mathrm{F})$,
we can use the asymptotic behavior of the Bessel functions
\begin{align}
J_l(\xi) &\sim \frac{(\sgn l)^l}{|\l|!}\left(\frac{\xi}{2}\right)^{|l|} \\
Y_l(\xi) &\sim -\frac{(\sgn l)^l (|l|-1)!}{\pi}\left(\frac{\xi}{2}\right)^{-|l|} \\
\frac{Y_l(\xi)}{J_l(\xi)} &\sim -\frac{(|l|!)^2}{\pi |l|} \left(\frac{2}{\xi}\right)^{2|l|}
\end{align}
for $|l| \gg \xi$.
The asymptotic behavior of the phase shift is given by
\begin{align}
& \cot\delta_j = -\frac{1+d}{2d} T_{l_\downarrow}(\xi_\mathrm{F}) + \frac{1-d}{2d} T_{l_\uparrow}(\xi_\mathrm{F}) \\
 &\simeq -\frac{1+d}{2d} \frac{(|l_\downarrow|!)^2}{\pi |l_\downarrow|} \left(\frac{2}{\xi_\mathrm{F}}\right)^{2|l_\downarrow|} + \frac{1-d}{2d} \frac{(|l_\uparrow|!)^2}{\pi |l_\uparrow|} \left(\frac{2}{\xi_\mathrm{F}}\right)^{2|l_\uparrow|} \\
 & \simeq
\begin{cases}
 -\frac{1+d}{2d} \frac{(|l_\downarrow|!)^2}{\pi |l_\downarrow|} \left(\frac{2}{\xi_\mathrm{F}}\right)^{2|l_\downarrow|} & (j>0)\\
\frac{1-d}{2d} \frac{(|l_\uparrow|!)^2}{\pi |l_\uparrow|} \left(\frac{2}{\xi_\mathrm{F}}\right)^{2|l_\uparrow|} & (j<0)
\end{cases} \\
 & = - \frac{d + \sgn j}{2 d} \frac{(\tilde{j}!)^2}{\pi \tilde{j}} \left(\frac{2}{\xi_\mathrm{F}}\right)^{2 \tilde{j}} + O(\xi_\mathrm{F}^{-(2\tilde{j}+2)}),
\end{align}
where $\tilde{j} = |j|+1/2$.
Thus we can see that the phase shift for large angular momentum is suppressed, as mentioned in the main text.

On the other hand,
for large momentum, or for a large skyrmion, i.e. $\xi_\mathrm{F} = k_\mathrm{F} R_\mathrm{S} \gg |j|$,
we can rely on the asymptotic behavior of the Bessel functions shown in Eqs.~(\ref{eq:asymptotic-J}) and (\ref{eq:asymptotic-Y}).
As a result, we obtain
\begin{align}
& \cot\delta_j  \sim -\frac{1+d}{2d} \tan\left[\xi_\mathrm{F} -\tfrac{\pi}{2}(l_\downarrow+\tfrac{1}{2})\right] \\
& \quad \quad \quad \quad \quad + \frac{1-d}{2d} \tan\left[\xi_\mathrm{F} -\tfrac{\pi}{2}(l_\uparrow+\tfrac{1}{2})\right] \nonumber \\
 & = \frac{1+d}{2d} \cot \left(\xi_\mathrm{F} - \tfrac{\pi}{2}j \right) +\frac{1-d}{2d} \tan \left(\xi_\mathrm{F} - \tfrac{\pi}{2}j \right) \\
 &= \frac{(1+d)\cos^2\left(\xi_\mathrm{F} - \tfrac{\pi}{2}j \right) + (1-d) \sin^2 \left(\xi_\mathrm{F} - \tfrac{\pi}{2}j \right)}{2d \sin\left(\xi_\mathrm{F} - \tfrac{\pi}{2}j \right) \cos\left(\xi_\mathrm{F} - \tfrac{\pi}{2}j \right)} \\
 &= \frac{1+d \cos\left(2\xi_\mathrm{F} - \pi j \right)}{d \sin\left(2\xi_\mathrm{F} - \pi j \right)} \\
 &= \frac{1 + (-1)^{j-1/2} d \cos(2\xi_\mathrm{F}-\tfrac{\pi}{2})}{(-1)^{j-1/2} d \sin(2\xi_\mathrm{F}-\tfrac{\pi}{2})} \\
 &= -\frac{(-1)^{j-1/2} +  d \sin(2\xi_\mathrm{F})}{d \cos(2\xi_\mathrm{F})},
\end{align}
where we have taken the leading order terms
and have neglected the subleading terms of $O(\xi_\mathrm{F}^{-1})$.

\section{Derivation of scattering amplitude}
Here we show the derivation process of the electron scattering amplitude by a single skyrmion,
whose final result is given in Eq.~(6).
As mentioned in the main text,
we consider here a case where an incident plane wave with the wave vector $\bfk$ and the energy $E=\sqrt{v_F^2 |\bfk|^2 +\Delta^2}$ is scattered by the skyrmion
and comes out as a cylindrical wave with the wave number $k=|\bfk|$.
The total wave function away from the skyrmion can be defined as the linear combination of the incident and outgoing waves,
\begin{align}
\Psi(\bfr) \sim \psi_{\bfk}^\mathrm{(in)}(\bfr) + \psi^\mathrm{(out)}(\bfr), \label{eq:Psi-scattering}
\end{align}
with the incident plane wave
\begin{align}
\psi_{\bfk}^\mathrm{(in)}(\bfr) &= e^{i\bfk\cdot\bfr} \vecv{-i\sin\tfrac{\zeta}{2}}{\cos\tfrac{\zeta}{2}}.
\end{align}
Here the outgoing wave component is defined by
\begin{align}
\psi^\mathrm{(out)}(\bfr) = \vecv{f_\uparrow(\phi)}{f_\downarrow(\phi)}\frac{e^{ik\rho}}{\sqrt{\rho}},
\end{align}
with $f_{\uparrow/\downarrow}(\phi)$ the scattering amplitude for each spin component.

As in the scattering theory of Schr\"{o}dinger electrons,
the scattering amplitude can be calculated in terms of the partial-wave decomposition.
Taking the $x$ axis to the direction of the incident wave vector $\bfk$,
the plane wave can be decomposed into partial waves as
\begin{align}
\psi_{\bfk}^\mathrm{(in)}(\bfr) &= e^{i k\rho \cos\phi} \vecv{-i\sin\tfrac{\zeta}{2}}{\cos\tfrac{\zeta}{2}} \\
 &= \sum_{l \in \mathbb{Z}} i^l J_l(k\rho) e^{i l\phi} \vecv{-i\sin\tfrac{\zeta}{2}}{\cos\tfrac{\zeta}{2}},
\end{align}
which shows the asymptotic behavior
\begin{align}
\psi_{\bfk}^\mathrm{(in)}(\bfr) \sim \sum_{l \in \mathbb{Z}} i^l e^{il\phi} \sqrt{\frac{2}{\pi k\rho}}\cos\left(k\rho -\frac{l\pi}{2}-\frac{\pi}{4}\right) \vecv{-i\sin\tfrac{\zeta}{2}}{\cos\tfrac{\zeta}{2}}.
\end{align}
For the scattering amplitude, we define the partial-wave decomposition
\begin{align}
f_\uparrow(\phi) &\equiv \sum_{l_\uparrow \in \mathbb{Z}} \sqrt{\frac{2}{\pi k}} \sin\frac{\zeta}{2} \ g_{l_\uparrow}^\uparrow e^{il_\uparrow \phi} \\
f_\downarrow(\phi) &\equiv \sum_{l_\downarrow \in \mathbb{Z}} \sqrt{\frac{2}{\pi k}} \cos\frac{\zeta}{2} \ g_{l_\downarrow}^\downarrow e^{il_\downarrow\phi}.
\end{align}

By comparing the both sides of  Eq.~(\ref{eq:Psi-scattering}) for each partial wave component in the limit $\rho \rightarrow \infty$,
we obtain a sequence of equations
\begin{align}
\alpha_{j} \cos\left[\xi_{j} -\delta_{j} \right] &= -i^{l_\uparrow +1} \cos \xi_{j} + e^{i \xi} g_{l_\uparrow}^\uparrow \label{eq:relation-up} \\
-\alpha_{j} \cos\left[\xi_{j} -\tfrac{\pi}{2} -\delta_{j} \right] &= i^{l_\downarrow} \cos\left[\xi_{j} -\tfrac{\pi}{2} \right] + e^{i\xi} g_{l_\downarrow}^\downarrow, \label{eq:relation-down}
\end{align}
where we use shorthand notations $\xi \equiv k\rho$, $\xi_{j} \equiv k\rho-(\pi/2)j$, and $l_{\uparrow/\downarrow} = j \mp 1/2$.
Since the second term in the right-hand side of each equation does not contribute to the incoming cylindrical wave $e^{-ik\rho}$,
the incoming parts of Eqs.~(\ref{eq:relation-up}) and (\ref{eq:relation-down}) give the same relation,
\begin{align}
\alpha_{j} e^{i[(\pi/2)j + \delta_{j}]} &= -e^{i(\pi/2)(j+1/2)} e^{i(\pi/2)j},
\end{align}
which fixes the value of the coefficient $\alpha_j$ as
\begin{align}
\alpha_{j} &= -e^{i [ (\pi/2)(j+1/2)-\delta_j ]}.
\end{align}
Substituting this relation to the outgoing $(e^{ik\rho})$ parts,
we obtain
\begin{align}
g_{l_\uparrow}^\uparrow &= \frac{\alpha_{j}}{2} e^{-i[(\pi/2)j + \delta_{j}]} + \frac{1}{2}e^{i(\pi/2)(l_\uparrow+1)} e^{-i(\pi/2)j} \\
 &= \frac{1}{2} e^{i(\pi/4)} \left[1- e^{-2i\delta_{j}}\right] \\
g_{l_\downarrow}^\downarrow &= -\frac{\alpha_{j}}{2} e^{-i[(\pi/2)(j+1) + \delta_{j}]} - \frac{1}{2}e^{i(\pi/2)l_\downarrow} e^{-i(\pi/2)(j+1)} \\
 &= \frac{i}{2} e^{i(\pi/4)} \left[1- e^{-2i\delta_{j}}\right].
\end{align}
Thus the total scattering amplitude is given by
\begin{align}
\vecv{f_\uparrow(\phi)}{f_\downarrow(\phi)} &= \frac{1}{2}e^{i(\pi/4)}\sqrt{\frac{2}{\pi k}} \label{eq:scattering-amplitude-phi} \\
 & \quad \times  \sum_{j \in \mathbb{Z}+1/2 } \left(1-e^{-2i \delta_{j}}\right) \vecv{-i \sin\tfrac{\zeta}{2} e^{i(j-1/2)\phi}}{\cos\tfrac{\zeta}{2} e^{i(j+1/2)\phi}}, \nonumber
\end{align}
which further reduces to Eq.~(6).

\section{Phase factor on transmitted plane wave at skyrmion boundary}

\begin{figure}[tbp]
\includegraphics[width=8cm]{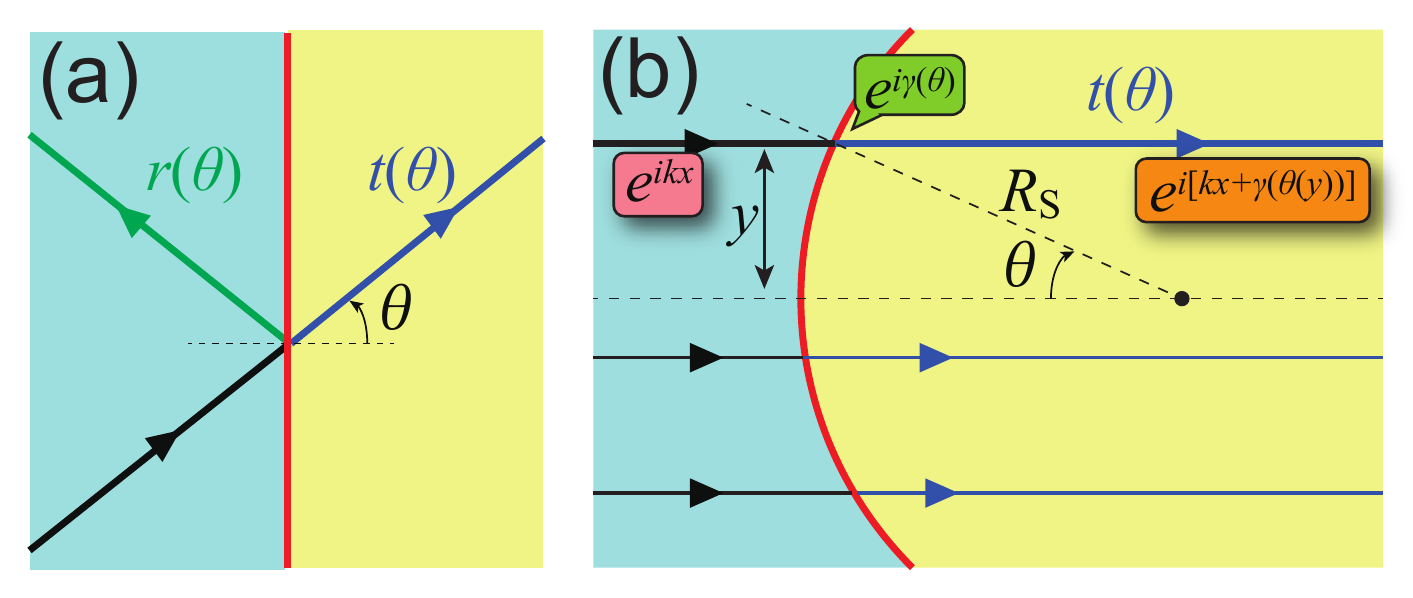} \\
\caption{Schematic pictures of the transmission of plane waves at the boundary
between two regions with opposite magnetizations.
(a) If a plane wave reaches a straight boundary,
the complex transmission rate $t$ and the reflection rate $r$ depend on the incident angle $\theta$.
(b) If the boundary has a curvature, the incident angle $\theta$ depends on the lateral position $y$,
leading to the position-dependent phase factor $e^{i\gamma(\theta(y))}$.
}
\label{fig:transmission}
\end{figure}

Here we give a detailed discussion on the geometric phase factor acquired at the skyrmion boundary,
which leads to the scattering skewness, as briefly mentioned in the main text.
We first start from the simplified geometry,
where the region with $n_z = +1$ and that with $n_z =-1$ is separated by the straight boundary at $x=0$,
as shown in Fig.~\ref{fig:transmission}(a).
If one takes a plane wave with the wave number $k$ and the incident angle $\theta$ as the incident wave,
its wave function is given by
\begin{align}
\psi_\mathrm{i}(\bfr) = e^{i k(x \cos\theta + y\sin\theta)} \vecv{-i \sin\frac{\zeta}{2} e^{-i\theta/2}}{\cos\frac{\zeta}{2}e^{i\theta/2}}.
\end{align}
Due to the translational symmetry in $y$-direction,
the $y$-component of the momentum is conserved through the transmission and reflection processes,
yielding the transmitted and reflected wave functions
\begin{align}
\psi_\mathrm{t}(\bfr) &= e^{i k(x \cos\theta + y\sin\theta)} \vecv{-i \cos\frac{\zeta}{2} e^{-i\theta/2}}{\sin\frac{\zeta}{2}e^{i\theta/2}} \\
\psi_\mathrm{r}(\bfr) &= e^{i k(-x \cos\theta + y\sin\theta)} \vecv{-i \sin\frac{\zeta}{2} e^{-i(\pi-\theta)/2}}{\cos\frac{\zeta}{2}e^{i(\pi-\theta)/2}},
\end{align}
respectively.
The complex transmission coefficient $t(\theta)$ and the reflection coefficient $r(\theta)$ should satisfy the boundary condition at $x=0$,
\begin{align}
\psi_\mathrm{i}(x=0;y) + r(\theta)\psi_\mathrm{r}(x=0;y) = t(\theta) \psi_\mathrm{t}(x=0;y).
\end{align}
Solving this equation, we obtain
\begin{align}
t(\theta) &= \frac{\cos\theta \sin\zeta}{\cos\theta - i \sin\theta \cos\zeta} \\
r(\theta) &= \frac{i \cos\zeta}{\cos\theta - i \sin\theta \cos\zeta}.
\end{align}
One can easily check that these factors satisfy the conservation of flux,
\begin{align}
|t(\theta)|^2 + |r(\theta)|^2 = 1.
\end{align}
The scattered wave,
namely the deviation of the transmitted wave from the incident wave, is characterized by the factor
\begin{align}
t(\theta) -1 &= \frac{\cos\theta (\sin\zeta-1) + i\sin\theta \cos\zeta}{\cos\theta - i \sin\theta \cos\zeta} \\
 &= -(1-\sin\zeta)\frac{1-i \cos\zeta(1-\sin\zeta)^{-1} \tan\theta}{1-i\cos\zeta\tan\theta},
\end{align}
which implies that the scattered wave acquires the geometric phase factor
\begin{align}
e^{i\gamma(\theta)} = \frac{t(\theta)-1}{|t(\theta)-1|}, \label{eq:phase-factor}
\end{align}
dependent on the incident angle $\theta$.

Next we set up a curved boundary, with the curvature radius $R_\mathrm{S}$, as shown in Fig.~\ref{fig:transmission}(b).
Here we focus on the behavior at a long wavelength so that the uncertainty in the position can be neglected.
Introducing an incident plane wave with the wave number $k$ in the $x$-direction,
its incident angle $\theta$ on the boundary depends on the lateral incident position $y$,
as $\theta(y) = \arcsin(y/R_\mathrm{S})$,
leading to the phase factor $e^{i\gamma(\theta(y))}$ for the scattered wave.
The phase $\gamma(\theta)$ defined by Eq.~(\ref{eq:phase-factor}) can be expanded around $\theta=0$ as
\begin{align}
\gamma(\theta) = \pi + \mu \theta +O(\theta^2) = \pi + \tilde{k} y + O(y^2),
\end{align}
where $\tilde{k} = \mu/R_\mathrm{S}$.
Noting the inequality
\begin{align}
\cos\zeta(1-\sin\zeta)^{-1} > \cos\zeta,
\end{align}
the factors $\mu$ and $\tilde{k}$ here are negative.
Therefore, the phase factor can be approximated as
\begin{align}
e^{i\gamma(\theta)} = - e^{i \tilde{k} y +O(y^2)},
\end{align}
which can be interpreted as the transverse wave component,
accounting for the scattering skewness.
Since the phase $\gamma(\theta)$ takes the opposite sign if the sign of $n_z(\bfr)$ is flipped,
the scattering skewness is sensitive to the skyrmion number.


\end{document}